# Carbon nanotubes heavy metal detection with stripping voltammetry: A review paper

Tingting Wang,* Wei Yue

Department of Chemistry, University of Cincinnati, Cincinnati, Ohio, 45221-0172, United States

*Corresponding author: wangt3@mail.uc.edu

## Abstract

The challenge of heavy metal detection for environmental, industrial and medical purposes has led to the development of many analytical techniques. Stripping voltammetry is a very sensitive electrochemical method and has been widely used for heavy metal detection. Carbon nanotubes, a well-studied carbon material with physical and chemical properties suited for electrode material is commonly employed for sensitive and selective metal detection in electrochemistry. This article reviews the recent (2011-2016) applications of carbon nanotubes as an electrode or electrode surface modifier for heavy metals detection with stripping voltammetry.

## The Necessities for Heavy Metal Detection

Heavy metals are generally referred as high atomic weight metallic chemical elements that could pose risks to the environment and living organisms.[1] Some of the metals that pose health risks are necessary for humans at trace levels but are toxic at higher levels (zinc, copper, manganese, nickel) while others are toxic with no known benefits (cadmium, mercury, lead). In addition, some toxic semi-metals such as arsenic and selenium are also considered to be as dangerous as heavy metals.[1,2] There are many adverse health effects associated with exposure to heavy metals at abnormal concentrations that can lead to various diseases. Exposure can result from industrial processes such as smelting, disposing of factory waste, as well as diet, medication, etc.[3] Heavy metal pollution is different from pollution by many organic species as metals do not decay into harmless compounds with time.[1,2] Due to the extensive industrial and agricultural use of heavy metals and the hazardous effect to human and other organisms, analytical approaches that can be used to monitor the presence of heavy metals in the environment and biological matrices such as blood, sweat, and urine are increasingly in demand. The complex relationships between the metal level in the body and toxic pathological effects have continuously driven for better analytical tools with improved sensitivity and selectivity for heavy metal detection.

## Stripping Voltammetry

The advantage of electroanalytical techniques, in general arises from their simplicity. Analysis systems utilizing electrochemistry are compact with few moving parts and are relatively simple to automate. In addition, they are often affected favorably by miniaturization due to the favorable signal to noise benefits associated with microelectrodes. Stripping voltammetry is a versatile electrochemical technique that has been used for a wide range of applications, and it has many advantages that make it suitable for both clinical and environmental applications.[4,5] The stripping voltammetry technique consists of two major steps: accumulation (pre-concentration) and stripping. During the accumulation step, the analyte is deposited at the working electrode surface. The accumulation of analytes through either an oxidation potential (cathodic stripping voltammetry, CSV) or reduction potential (anodic stripping voltammetry, ASV) or the adsorption to the electrode surface (adsorptive stripping voltammetry, AdSV) is followed by the measurement of current during an opposite potential scan that reversed the redox process used or deposition.[6] ASV is mostly used in trace metal analysis and multiple metal ions can be measured simultaneously in various matrices.[6] In this case, by applying a potential significantly negative to reduce metal ion ($M^{n+}$ to M) is followed by the re-oxidation of M to $M^{n+}$, and the peak current, which is proportional to the concentration of $M^{n+}$ in the solution, is recorded simultaneously. In AdSV, often a suitable ion coupling/extraction reagent(s) is present on the electrode which can bind ions with the electrode. These processes result in the analyte pre-concentration that provides the ability to obtain nanomolar detection limit.[6] In most cases, only a small fraction of the total analyte in the sample is deposited because complete deposition of all the analyte would be time-consuming and is usually unnecessary to reach the required limit of detection. Since the deposition is not exhaustive, it is important to reproducibly control the fraction of analyte in the sample that is deposited. To do this, electrode surface area, deposition time, and rate of convective mass transport must be carefully optimized and controlled throughout the experiment. Deposition time can vary widely depending on the analyte concentration, the type of electrode, and the stripping technique used. The stripping signal is proportional to the total amount of analyte deposited, so less concentrated solutions require longer deposition times in order to achieve adequate stripping peaks.[4,6,7]

## Carbon Nanotubes and the Electroactivity of Carbon Nanotubes

The traditional electrodes used for trace metal analysis are the dropping mercury (DME), hanging mercury drop electrode (HMDE) and mercury film electrode (MFE). Mercury-based electrodes have been used for years because of their reproducible electrode surface, good negative potential window, ability to dissolve the deposited metals and excellent performance with stripping voltammetry.[8] But due to their toxicity and the cost associated with their disposal, mercury-based electrodes are no longer the best choice for sustainable applications, and alternative electrodes have been heavily studied.[9–11] As different nanoparticles been invented and characterized, many of them are explored in various applications including electrochemistry.[12–16] Since they were first discovered, carbon nanotubes (CNTs) have been attracting increased attention thanks to their excellent electrochemical properties, including wide potential window, fast electron transfer rate and large surface area.[17–22] Applications with CNTs have a wide range of use, including in biomedicine, nanoelectronics, environmental engineering and electrochemical sensing.[23,24] In this review, we focus on selected applications of CNTs for heavy metal detection by stripping voltammetry from the past five years (2011-2016).

## Fabrication of CNT electrodes

Generally, CNTs have two classes: single-walled carbon nanotubes (SWCNTs) and multi-walled carbon nanotubes (MWCNTs). SWCNTs consist of a simple geometry with a rolled single graphene sheet which has a diameter within the range of 0.4-2 nm and up to 20 cm in length.[25] MWCNTs consist of several concentric tubes fitted into each other with a diameter up to 100 nm.[26,27] Depending on the desired properties of the CNTs, different methods have been developed to synthesize CNTs. There are three main methods for the synthesis of CNTs: arc discharge, laser ablation, and chemical vapor deposition (CVD).[28] Arc discharge and laser ablation belong to the high temperature (above 3000 K) and short time (μs-ms) synthesis techniques.[28] Prasek et al. have provided a thorough discussion of the synthesis of MWCNTs and SWCNTs using arc discharge and laser ablation.[27] Although arc discharge and laser ablation produce high-quality nanotubes with superior straightness and crystallinity, however, the disadvantage of a very energy intensive procedure that requires a large amount of solid carbon/graphite as the substrate has limited their application for large scale production.[28–30] Compared to the higher temperature synthesis techniques, lower temperature synthesis methods of CVD provide better control of CNT orientation, nanotube length, diameter, alignment, and yield, making it the most prevalent technique for CNT growth nowadays.[27] In CVD, CNTs can be synthesized using metal or non-metal substrates and the reaction rate of CNT formation and site density can be controlled with the application of electron-beam lithography.[29] Figure 1 is a typical schematic of a CVD reactor: hydrocarbon precursor is carried by argon saturated with water, which serves as an oxidizing reagent to remove amorphous carbon, into the furnace.[31] Patterned metal catalysts are placed on the sample holder surface.[31,32]

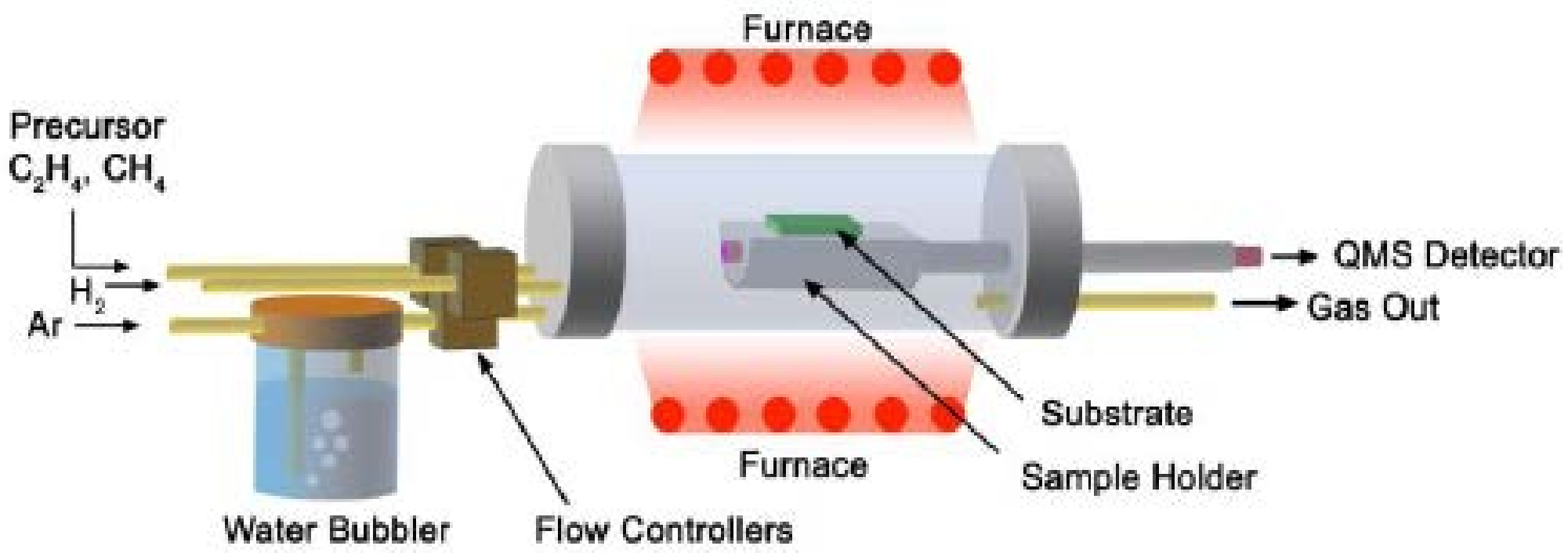

Fig 1. Typical schematic of CVD reactor for CNTs growth.[31] (Reprint permission was acquired from the Journal of the University of Chemical Technology and Metallurgy).

CNTs are formed by the decomposition of hydrocarbon precursors with a continuous flow of gas on patterned metal catalyst sites.[31,32] Previously it was reported that CNTs have been grown with controllable length on iron-silicon substrates with ethylene and nickel-glass substrates with acetylene.[33] During the growth, neighboring CNTs interact with each other via Van der Waals forces to form high-density CNT bundles with rigidity. In general, there are two steps for catalytic CNT formation. First, a meta-stable carbide is formed at the catalyst surface, resulting in the formation of an amorphous carbon rod in a fast step. This is followed by graphitization of the carbon rod which allows the slow formation of the hollow tubes and the termination of growth.[34]

There are two common growth mechanisms of CNTs on the metal catalysts.[29] (a) In tip growth model where the catalysts-substrate interaction is weak, the precursor diffuses through the metal catalyst and decomposes on the catalyst surface. As long as there is an open space for the precursor to decompose, CNTs continue to grow and push the catalyst out from the substrate until the catalyst is completely covered with CNTs. (b) In base growth mode where the catalyst-substrate interaction is strong, the precursor diffuses over the surface of the catalyst and as the CNT precipitate fails to push the catalyst from the substrate, it continues to emerge out from the top of the catalyst.[29] The precursors that are most commonly used for growing CNTs are methane, ethylene, benzene, xylene and carbon monoxide. Popularly used metal catalysts are transition metals (Fe, Co, Ni) and some less commonly used metals (Cu, Pt, Pd, Mn, Mo, Cr, Sn, Au, Mg, Al) have also been reported for the CVD process. Two previous reviews have comprehensively discussed the proper selection of precursors and catalysts for CVD.[29,32]

The pursuit of improvements in the field of CNT fabrication led to the development of the non-metal catalyzed system. Liu et al. have reported that a non-metallic substrate ($SiO_2$ particles) acts as a catalyst for CNT growth using CVD of ethanol without catalysts. The annealing of $SiO_2$ in hydrogen at high temperature leads to the formation of defects at $SiO_2$ surfaces, which provide nucleation sites for the CNT growth. [35]

Another process used to produce aligned CNTs in a non-metallic catalytic system is the sublimation and decomposition of SiC at high temperature in a vacuum, which was first reported in 1997 by Kusunoki.[36] More recently, CNT derivative solid carbon nanorods (SCNRs) fabricated by Carbo-Thermal Carbide Conversion (CTCC) has been reported.[37] CTCC is a chemical process for producing carbon nanostructures from solid phase carbide source materials. The commonly used substrate is carbide ceramics, such as SiC in the states of single crystal, polycrystalline or amorphous. In the CTCC process, the metal or metalloid impurities from a carbide material react with reducing gasses ($H_2O$, $C_xO_y$, air or a mixture of these gasses) and form gaseous compounds at a high processing temperature. As a result, the metal/metalloid species can be selectively removed leaving only carbon species. CTCC-grown materials are also referred to as Solid Carbon Nanorods (SCNRs) since they do not have the traditional hollow core but instead have a solid core structure inside. SCNRs are similar in appearance, physical and chemical properties to CVD grown CNTs, with concentric spacing on the order of interplanar spacing of graphite, approximately 0.4 nm.[38] The spectroscopic and electrochemical properties of single-walled carbon nanotube material produced by the CTCC process have been characterized.[39] Metallic catalyst-free CNTs (MCFCNTs) synthesized via a solid-phase growth mechanism have also been reported, [40,41]

**Heavy Metal Detection with CNTs using Stripping Voltammetry**

Analytical techniques of inductively coupled plasma-mass spectrometry, atomic absorption spectroscopy, atomic fluorescence spectroscopy and electrochemistry are commonly used for heavy metal detection.[2] Stripping voltammetry, as one of the most sensitive electrochemical techniques for metal determination, has been reported extensively for detection of Pb, Cd, Hg, As, Zn, Cu, Ag, Cr, Ni, and Ti, etc.[42–52] CNTs, as well as other carbon nanostructures, offer distinctive advantages of large surface area, fast electron transfer rate, high sensitivity and low cost hence have been widely researched for use in electrochemical metal ion sensing.[1,2] Previous reviews had discussed some representative stripping voltammetric detection of heavy metal ions with carbon nanostructures, and this review focuses on the most recent discovery and improvement of using CNTs in heavy metal ion sensing.[1,2]

Many forms of CNTs have been explored for use in stripping voltammetry. Free standing CNTs produced by CTCC or CVD, when used without other modifications are known as “direct analysis” for heavy metals. Electrodes cast with dispersed CNTs with modifications by other chemical reagents are “indirect analysis”, which still dominate in stripping voltammetric heavy metal sensing.

## 1. Bare CNTs electrodes for direct analysis

The direct use of CNTs as an electrode material, such as CNTs tower, CNTs array, and CNTs thread, provides compelling simplicity in electrode fabricating compared to modified electrodes discussed in the next section. The simultaneous detection of multiple heavy metal ions of Pb, Cd, Zn, Cu using CNTs tower has demonstrated that even with a simple form of the CNTs, a good resolution for all 4 metal species can be achieved.[53] A catalysts-free CNTs, where nanotubes were directly grown on the non-metallic SiC substrate, also showed nanomolar detection limit for Pb, Cd, and Zn.[41] A miniaturized CNTs thread microelectrode has achieved low detection limits for Pb, Cd, Zn, and Cu and showed potential for clinical and medical applications.[54] Compared to conventional electrodes such as hanging mercury drop electrode (HMDE) and Bismuth (Bi) film electrode, stand alone CNTs offer the ability to detect multiple ions simultaneously and similar detection limits but with narrower linear ranges. (Table-1). But more importantly, CNTs electrodes do not pose the toxicity as mercury electrode, so they are much more friendly to the environment. Other bare electrodes such as glassy carbon (GC) and Platinum (Pt) are rarely used for direct analysis of heavy metal ions due to their poor sensitivity, but they are commonly used with CNTs modifications and will be discussed below. Bi film electrode, which was introduced almost two decades ago, is often compared with mercury electrode. And like CNTs, bismuth electrode is environment friendly with negligible toxicity, however, on the other hand, unlike the standing alone CNTs, Bi film is required to be deposited on a supporting electrode, such as GC thus is rarely used alone. Bi is also found useful when deposited together with CNTs to improve the sensitivity to the heavy metal ions. [10,55]

Overall, although easy to work with, the simple forms of CNTs electrode only represent a very small amount of all the CNTs-based heavy metal ion sensors probably due to the fact that they are mainly synthesized with unconventional approaches and also are very likely to encounter issues such as selectivity especially when no other modifications were made to improve the specificity toward the analyte, in the case of interfering ions are present in complex samples.

Table-1 Direct analysis of heavy metal ions with CNTs electrodes in comparison with HMDE and Bi film electrode

| Electrode | Target ion | Simultaneous Detection? | Real sample | Linear range | Detection limit | Year published |
|---|---|---|---|---|---|---|
| MWCNTs tower | Pb, Cd, Cu, Zn | Yes | - | - | 2.5, 2.8, 2.8, 4.4 ppb | 2011[53] |
| Metal-catalyst-free CNTs | Pb, Cd, Zn | Yes | - | 0.062-1.7 ppm<br>0.056-0.56 ppm<br>0.033-0.46 ppm | 2.7, 3.6, 3,3 ppb | 2012[41] |
| SWCNTs film | Pb, Cd | Yes | - | 0.033-0.228 ppm | 0.7, 0.8 ppb | 2012 [56] |
| MWCNTs thread | Zn | - | - | 0.065- 0.92 ppm | 0.092 ppb | 2013 [48] |
| MWCNTs thread | Pb, Cd, Cu, Zn | Yes | - | 0.21-0.83 ppm<br>0.17-0.51 ppm<br>0.032-0.22 ppm<br>0.20-0.59 ppm | 0.31, 0.21, 0.017, 0.091 ppb | 2014 [54] |
| CNTs fiber | Pb | - | - | - | 0.36 ppb | 2016 [57] |
| HMDE | Pb, Cd, Cu, Zn | Yes | Algae | 0.0005-0.76 ppm<br>0.0005-1 ppm<br>0.0015-0.52 ppm<br>0.005-0.76 ppm | 0.48, 0.4, 1.3, 3.2 ppb | 1998 [58] |
| SPE-Bi | Pb | - | - | 0.01-0.1 ppm | 0.3 ppb | 2001 [59] |

SPE: screen-printed carbon electrode

## 2. Modified CNT electrodes

In order to improve the performance of CNTs electrodes, a variety of strategies for modifying the electrode surface have been explored. These are generally designed to provide better selectivity or to lower the limits of detection.

### 2.1. Covalent modification: –COOH, -$NH_2$ and -SH

Pristine CNTs are often pretreated by exposure to acids such as $HNO_3$, $H_2SO_4$, HCl or a mixture of them to eliminate amorphous carbon and metal catalyst impurities and to introduce carboxyl and carboxylate groups at defects and the end of the CNTs surface. Functionalization by carboxylic groups can improve the conductivity of the electrode and cause electrostatic interaction between the CNTs electrode and target cations that improve the electrochemical signal.[60–65] Functionalization with amino groups (-$NH_2$) and thiol groups (-SH) has also been used to preconcentrate metals on the electrode surface by coordination through the interaction between electron-rich ligands and electron-deficient heavy metal ions.[66–68] An $NH_3$ plasma treated MWCNTs that was used as a sorbent to extract and preconcentrate Cu, Cd, Zn and Hg achieved sub-nanomolar detection limit with stripping voltammetry. The low detection limit raised from the improved faradic current to the capacitive current ratio attributed to the high electron conduction of MWCNTs and the good adsorptive capacity of $NH_3$.[66] (Figure 2)

A simultaneous thiol and N-doped functionalization have made the MWCNTs a good sensor for Pb and Cd. The thiol group can improve the metal ions selectivity and enrichment ability due to coordination of the metal ions by the surface thiol groups.[68] Similarly, amino acids are useful for CNTs modification because of their richness in both $–NH_3^+$ and –COOH groups. [45,52,69,70] An optimized leucine/MWCNTs sensor for As showed a detection limit as low as 1.67 nM (0.125 ppb), and a glutamine modified MWCNTs sensor reached a detection limit of 37.3 nM (2.79 ppb) for As. Both sensors posed the detection limit well below the EPA requirement for As in drinking water (~135 nM, 10 ppb) and were demonstrated effective in determining As in waste water.[37,44, 62] Table 2 summarizes the recent publications of covalently modified CNTs for heavy metal detection.

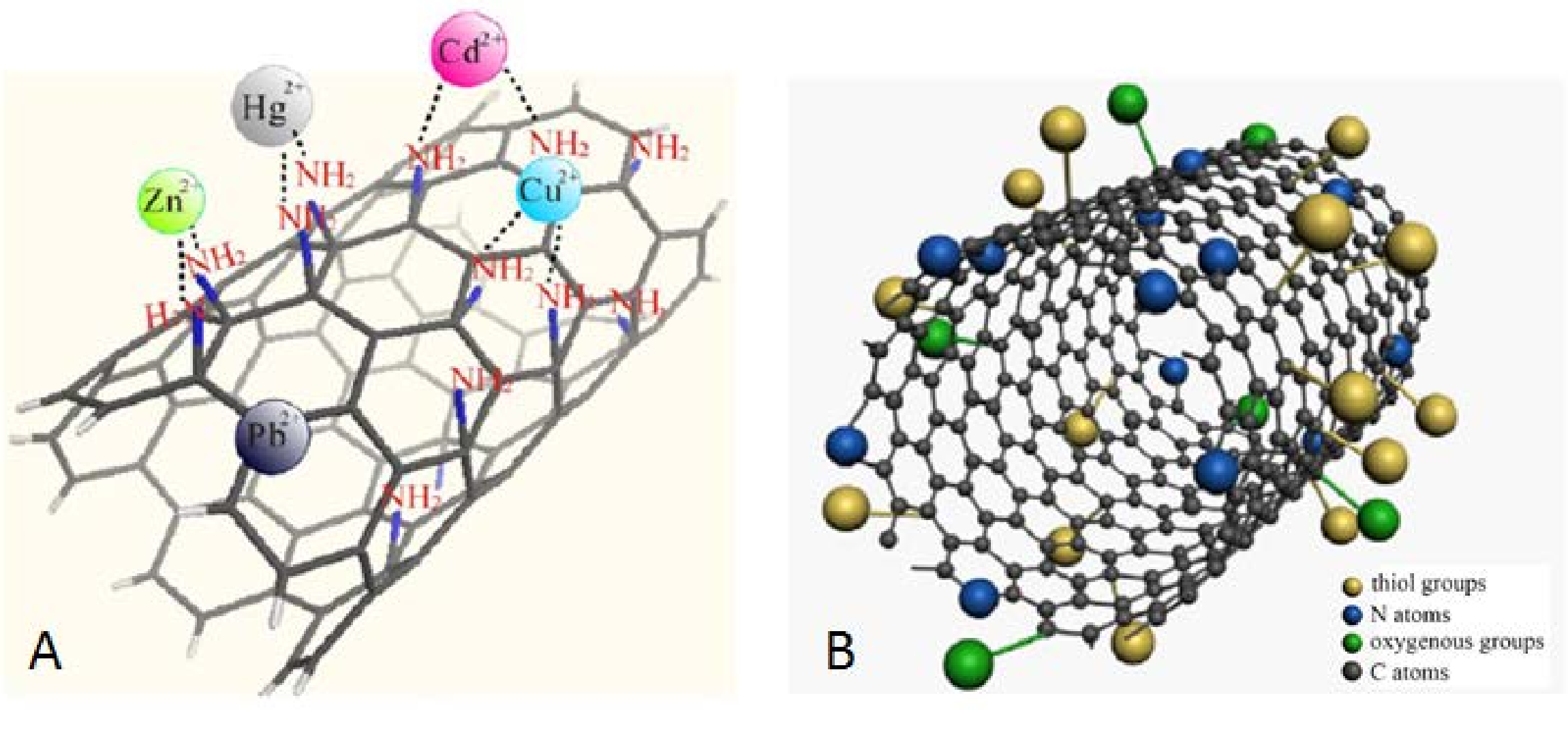

Fig. 2 (A) Schematic representation of possible interactions between $Pb^{2+}$, $Cd^{2+}$, $Cu^{2+}$, $Cd^{2+}$, $Hg^{2+}$ and $NH_3$ plasma treated CNT.[66] (B) Schematic of modification of CNT with multiple treatments.[68] (Reprint permission was acquired from Elsevier)

Table 2 Representative covalent modification of CNTs electrode for heavy metal detection

| Electrode | Covalent Modification | Target ion | Simultaneous Detection? | Real sample | Linear range | Detection limit | Year published |
|---|---|---|---|---|---|---|---|
| (SWCNTs/Hg/Bi)-GC | $-COO^-$, -OH | Pb, Cd, Zn | Yes | River water | 0.005-1.1 ppb<br>0.5-11 ppb<br>10-130 ppb | 0.12, 0.076, 0.23 ppb | 2011[60] |
| MWCNTs/Nafion/PDMcT/Bi-GC | -SH | Pb, Cd | Yes | Tap and spring water | 0.1-22 ppb<br>0.05-20 ppb | 0.03 ppb<br>0.05 ppb | 2011 [61] |
| MWCNTs/5-Br-PADAP/GC | $-COO^-$, -OH | Pb | - | Spring, drinking, waste water | 0.9-114 ppb | 0.1 ppb | 2011 [72] |
| MWCNTs/Nafion-Pt | Glutamine | As | - | Waste water | - | 2.7 ppb | 2011 [52] |
| MWCNTs/Bi/Nafion-SPE | $-COO^-$, -OH | Pb, Cd, Zn | Yes | - | Multiple* | Multiple* | 2012 [73] |
| MWCNT-GC | $-NH_2$ | Cd, Zn, Cu, Hg | Yes | Reservoir water | - | 0.003, 0.021, 0.014, 0.029 ppb | 2012 [66] |
| MWCNTs/Nafion-Pt | Leucine | As | - | Waste water | 1.5-150 ppb | 0.13 ppb | 2012 [45] |
| SWCNTs-AuNP-GC | L-Cystein | Cu | - | River water | 0.0060-8.9 ppb | 0.0012 ppb | 2013 [69] |
| MWCNTs/CTS/GA-SPE | $-COO^-$, -OH | Pb | - | Natural water | 0.099-2 ppb | 11.8 ppb | 2014 [74] |
| MWCNTs-GC | Polyhistidine | Cu | - | Tap and ground water | 0.032-7.9 ppm | 4.8 ppb | 2015 [70] |
| CNTs/Nafion-GC | - $NH_2$, $-COO^-$, -OH | Pb, Cd | Yes | Tap and ground water | 0.0021-14.5 ppm<br>0.011-11.24 ppm | 0.21 ppb<br>0.56 ppb | 2015 [67] |
| MWCNTs/Bi-GC | -SH, $-NH_2$, $-COO^-$, -OH | Pb, Cd | Yes | - | 2-50 ppb | 0.3 ppb<br>0.4 ppb | 2016 [68] |

PDMcT: poly(2,5-dimercapto-1,3,4-thiadiazole), formed –SH groups under reduction potential.
5-Br-PADAP: 2-(5-bromo-2-pyridylazo)-5-diethylaminophenol; GC: Glassy carbon; SPE: Screen printed electrode; CTS: Chitosan; GA: Glutaraldehyde; Bi: Bismuth

## 2. Non-covalent modification

Compared to direct analysis of heavy metal ions using an electrode consisting only of bare CNTs, electrodes of GC, Pt, Au, or carbon paste electrode (CPE) modified by CNTs to transform the surface into a CNT electrode and used as ion sensors are more common. The combination of CNTs with these platforms can improve the flexibility of CNTs, as most of the commercial CNTs is in its power form. Also, with the combination, electrodes such as GC, CPE and SPE can provide a solid and conductive surface for CNT and make the electrode much easier to reproduce. But the main challenge for developing this type of CNT modified electrode is that CNTs do not dissolve or disperse well in most of the common organic and inorganic solvents since the high

Van der Waals force causes CNTs to aggregate into bundles.[75,76] Many approaches were proposed to dissolve CNTs through non-covalent interactions, mainly through the adsorption ability of organic modifiers on the CNTs surface or π-π conjugation between CNTs and aromatic molecules.[77] A very recent review has compared these strategies in detail.[78]

## 2.1 CNT dispersion

Surfactants can successfully solubilize CNTs in aqueous solution by sheathing the CNTs surface with hemimicelles.[79,80] Figure 3 shows a schematic representation of how surfactants can adsorb onto the nanotube surface. In general, alkyl groups of surfactants lie on the CNTs surface parallel to the tube axis. Surfactants with benzene-rings have better dispersing ability due to the π-π stacking interaction between surfactant and CNTs, which can increase the binding surface significantly. In addition, both headgroups and chain length on the surfactant can change the dispersing ability.[79] Dodecylbenzene sulfonate (NaDBS) and dihexadecyl phosphate have been used for CNTs dispersion for Zn, Cd and insulin detection.[81,82]

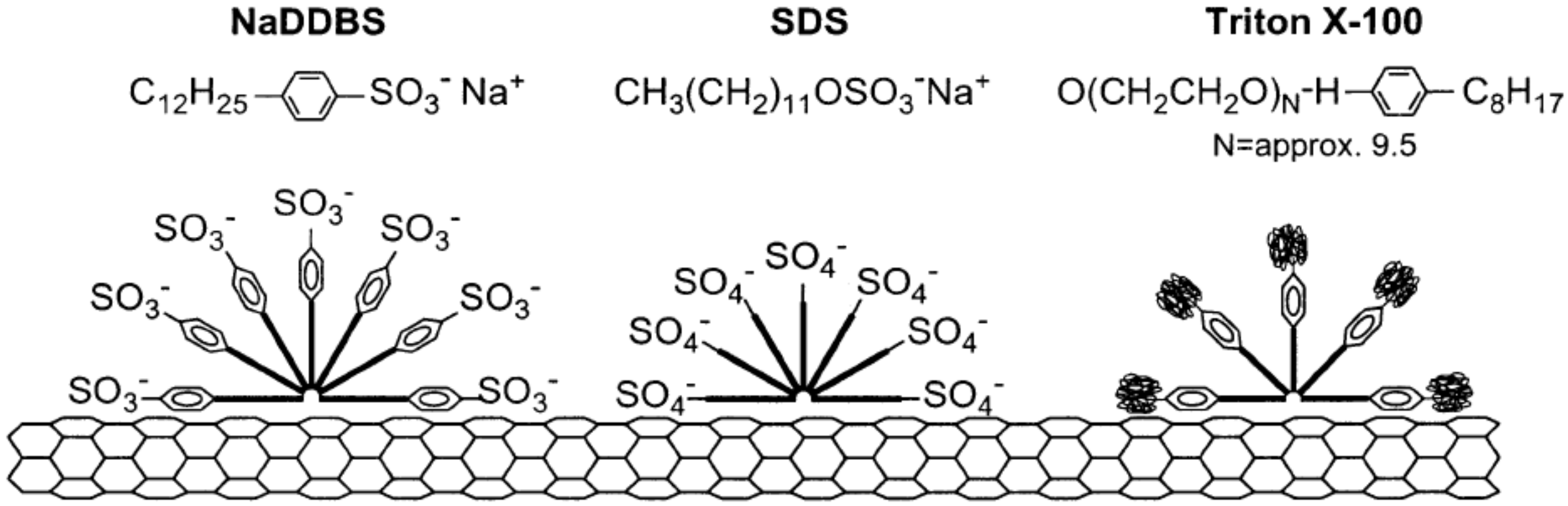

Fig. 3 Schematic representation of how surfactant can absorb onto CNT surface. [79] (Reprint permission was acquired from ACS publications)

Another approach for increasing CNTs dispersion in water is to use a high molecular weight polymer to wrap around the surface of CNTs through interactions including π-π, CH-π and

cation-π.[78]

In heavy metal ion detection, due to its advantages of thermal stability, chemical inertness, mechanical strength, and resistance to fouling, Nafion has been widely used to disperse CNTs and modify electrodes in electrochemistry.[83,84] One example is found in a recent study where a GC electrode was modified with Nafion-MWCNT and used as a $Eu^{3+}$ sensor.[85] The mechanism of MWCNT concentration dependent effects on the electrochemical signal is: without the presence of MWCNTs, $Eu^{3+}$ concentration is enhanced at the GC electrode surface through ionic attraction with Nafion, where electron transfer is achieved by diffusion of $Eu^{3+}$ to the GC electrode surface. When low concentrations of MWCNTs are present, few interconnections are established among individual nanotubes. Thus, no significant increase in the microscopic surface area of the electrode surface is caused by the presence of MWCNTs, and the electron transfer rate is comparable to that of Nafion alone. Thus, the same diffusion distance across the film and charge-transfer mechanism occur as with only Nafion present. When MWCNT concentrations increase to a **"**critical point**"**, the individual nanotubes begin to connect and significantly increase the microscopic surface area of the electrode. This effectively extends the electrode surface area into the film, enabling electron transfer to now occur within the Nafion film which reduces the diffusion distance. Instead of happening only at the GC electrode surface, the reduction of $Eu^{3+}$ to $Eu^{2+}$ now also occurs at both the underlying GC electrode surface and the MWCNT matrix within the film.[85] Other polymers used to disperse CNT but less frequently used in metal ion sensors can be found in a recent review.[78]

### 2.2 Improve selectivity

To improve the selectivity for heavy metal ions with CNT-based electrodes, chelating reagents such as ionophores and Schiff bases are commonly used as they can exclusively interact with metals ions. Ionophores, also known as ion carriers, offer defined cavities for specific metal ion incorporation to extract specific ions from aqueous solutions into hydrophobic membranes.[86] They are unique, sensitive and robust. However, ionophores are normally non-conductive and this property has restrained their wide application to amperometric analysis.[87] With the help of conductive polymer films along with CNTs, ionophores can form a matrix with excellent sensor conductivity and selectivity.[86–88] In 2012, a functionalization of MWCNT with thiacalixarene (an ion selective material/ionophore) was demonstrated to overcome the disadvantages of lack of conductivity of thiacalixarene and insufficient selectivity of CNTs to successfully develop a $Pb^{2+}$ sensor.[89] Similarly, a phytate-CNT-based $Cu^{2+}$ sensor was produced based on the stability of the phytate-metal complex which follows an order of Cu(II)>Zn(II)>Co(II)>Mn(II)>Fe(II)>Ca(II), and allowed the use of phytate to enrich and extract $Cu^{2+}$ with little interference from other ions.[90]

Schiff bases, another kind of frequently used reagent for forming complexes with metal ions, can also couple with CNTs for metal ion detection, such as $Pb^{2+}$ and $Hg^{2+}$, $Cd^{2+}$, $Cu^{2+}$.[91–94] Other modifications with similar ideas for complexing target heavy metal ions with one or multiple complexing reagents for extracting are summarized in Table-3. However, even with the assistance of ion selective chemicals (shown in Table-3), the sensors can still encounter sufficient interferences, especially in more complex samples such as in blood and urine, making the analysis inaccurate.[95] The development and exploration of more selective and sensitive ion complexing reagents for specific sample analysis are still needed and requires more studies.

Table 3. Representative modifications of CNT for improved selectivity

| Electrode | Modifications | Target ion | Simultaneous Detection? | Real sample | Linear range | Detection limit | Year published |
|---|---|---|---|---|---|---|---|
| MWCNT-GC | Nafion | Pb, Cd | Yes | River, tap and drinking water | - | 0.1 ppb<br>0.15 ppb | 2011 [62] |
| MWCNT-GC | PANI | Pb | - | - | - | - | 2011 [63] |
| MWCNT-SPE | Chitosan/mineral oil | Cd, Hg | - | Human urine/natural/ Waste water | 6.6-168.6 ppb<br>1.3-16.6 ppb | 1.1 ppb<br>0.48 ppb | 2011 [64] |
| MWCNT-CPE | Schiff base | Pb, Hg | Yes | Sea water/Shrimp/ human teeth | 0.41-145.0 ppb<br>0.40-140.4 ppb | 0.12 ppb<br>0.18 ppb | 2012 [91] |
| MWCNT-GC | Ionophore | Pb | - | River and tap water | 0.3-50 ppb | 0.1 ppb | 2012 [86] |
| MWCNT-CPE | Cu-ion-imprinted polymer | Cu | - | tap/well/ dam/riverwater/ human hair | 2-20 ppb | 0.34 ppb | 2012 [47] |
| MWCNT-GC (figure) | Thiacalixarene | Pb | - | - | 0.04-2.1 ppb | 8.3 ppt | 2012 [89] |
| MWCNT-GC | Cupferron/β-naphthol/DMF | Cd | - | Natural/waste water | 0.005-1.8 ppb<br>1.80-159.6 ppb | 1.80 ppt | 2012 [96] |
| MWCNT-CPE | PPh3 | Pb, Cd, Hg | Yes | waste water/gasoline/dry fish sample | 0.021-31.1 ppb<br>0.011-16.9 ppb<br>0.020-30.1 ppb | 0.012 ppb<br>0.010 ppb<br>0.015 ppb | 2013 [97] |
| MWCNT-CPE | Schiff base | Pb, Cd | Yes | Tap/waste water/shrimp/fish/ humanhair/tobacco | 0.4-1100 ppb<br>1-1200 ppb | 0.25 ppb<br>0.74 ppb | 2013 [92] |
| MWCNT-CPE | Schiff base | Ag | - | Tap/well/waste water | 0.5-235 ppb | 0.08 ppb | 2013 [98] |
| MWCNT-GC | $Cu_3(BTC)_2$/DMF | Pb | - | Tap/well/lake water | 0.21-10.4 ppb | 0.16 ppb | 2013 [99] |
| MWCNT-GC | Nafion/Benzo-18-crown-6 | Pb | - | Tap water | 0.21-6.2 ppb | 0.21 ppb | 2013 [100] |
| MWCNT-Au | Nafion | Zn | - | Bovine serum | 32.7-457.7 ppb | 3.47 ppb | 2013 [101] |
| MWCNT-CPE | Schiff base | Cd | - | Various water samples/human hair/milk powder/fungi/tobacco | 0.2-23 ppb | 0.08 ppb | 2014 [93] |
| MWCNT-ITO | Phytate | Cu | - | River water | 0.63-63.6 ppb | 158.9 ppb | 2014 [90] |
| MWCNT-GC | Nafion | Eu | - | - | 0.15-15.2 ppb | 0.056 ppb | 2014 [85] |
| MWCNT-CPE | Hydroxyapatite/ [dmIm][PF6]IL | Pb | - | Oil field water | 0.27-2072 ppb | 0.004 ppb | 2014 [102] |
| MWCNT-Pt | Nafion/ P1,5-DAN | Pb, Cd | Yes | River water | 4-150 ppb | 2.1 ppb<br>3.2 ppb | 2015 [103] |
| MWCNT-CPE | Schiff base | Cu | - | River water | 0.09-340 ppb | 0.01 ppb | 2015 [94] |
| CNF-GC | Nafion | Pb, Cd | Yes | - | - | 0.19 ppb<br>0.31 ppb | 2015 [104] |
| MWCNT- pyrolytic graphite | Salicylaldehyde azine/Nafion | Cu | - | River water | 0.32-19.1 ppb | 0.064 ppb | 2015 [105] |
| MWCNT-CPE | ACABP | Ag | - | Tap/well/waste water/tea leafs | 0.5-270 ppb | 0.079 ppb | 2015 [106] |
| MWCNT-GC | Poly(PCV)/Bi | Pb, Cd | Yes | Tap/well water | 1-200 ppb<br>1-300 ppb | 0.4 ppb<br>0.2 ppb | 2015 [107] |
| MWCNT-GC | BiOCl/Nafion | Pb, Cd | Yes | Sediment pore water | 5-50 ppb | 0.57 ppb<br>1.2 ppb | 2015[108] |
| SWCNT-Au | PABS/ DMAET | Hg | - | Waste water | 20-250 ppb | 12.0 ppb | 2016 [109] |
| MWCNT-CPE | iodoquinol | Cu | - | Oil and tap water | 0.64-317.8 ppb | 0.32 ppb | 2016 [95] |
| MWCNT-GC | β-CD/Nafion/Bi | Pb, Cd | Yes | Soil | 1-100 ppb | 0.13 ppb<br>0.21 ppb | 2016 [110] |
| GN-hyxCNTs-CPE | - | Cu | - | Pond water | 1.3-705.4 ppb | 0.6 ppb | 2016 [111] |
| | | | | | 1.98-7.06 ppm | | |
| AuNPs-CNFs-GCE | Isopropanol/Nafion | Pb, Cd, Cu | Yes | - | 20.7-207.2 ppb<br>11.2-112.4 ppb<br>6.4-63.5 ppb | 20.7 ppb<br>11.2 ppb<br>6.4 ppb | 2016 [112] |

PANI: polyanilineβ-CD: β-cyclodextrin; DMF: dimethylformamide; ; PPh3: Triphenyl phosphine; [dmIm][PF6]IL: 1-dodecyl-3-methylimidazolium hexafluoro phosphate ionic liquid; P1,5-DAN (poly(1,5-diaminonaphthalene)); ACABP: N_(4_{4_[(anilinocarbothioyl)amino]benzyl}phenyl)_N_phenylthiourea; PABS: poly (m-amino benzene sulfonic acid); DMAET: dimethyl amino ethane thiol

## 2.3 Other types of CNT-based electrodes

Besides the above frequently used CNT modifications and electrode platforms, disposable CNT-based sensors and flow through systems have also been studied for metal ion analysis. The screen-printed electrodes (SPE) are cheap and often used as disposable sensors. An MWCNT and Nafion modified SPE with in situ plated bismuth film showed ultra-sensitivity and simultaneous determination capability for $Zn^{2+}$, $Cd^{2+}$, and $Pb^{2+}$.[113] A glutathione (-SH rich peptide) modified SPE-carbon nanofiber electrode showed good performance toward $Pb^{2+}$ and $Cd^{2+}$.[114]

A flow-through system composed of CNT film deposited on polytetrafluoroethylene (PTFE) membrane was constructed and $Cu^{2+}$ was used to characterize the system as an example. Unlike the smaller devices, this type of system can not only detect metal ions but also be used in large scale electrochemical treatment for removing heavy metals from water.[115]

Hybrid composite electrodes of CNTs with other types of carbon or metal oxides have also received attention for improved heavy metal detections. Gold nanoparticles (AuNPs) deposited onto SWCNT film by cyclic voltammetry displayed a much higher sensitivity toward Pb and Cu compared with SWCNTs alone.[116] Besides, graphene oxide (GO) could be used to modify MWCNTs to form a 3D structure that showed an excellent solubility due to the hydrophilicity of GO. The sensor made from this 3D structure was successfully used to detect Pb and Cd.[117] Table-4 summarizes other composite CNT sensors for heavy metal ions.

Table 4. CNT modification with other electrode materials

| Electrode | Modifications | Target ion | Simultaneous Detection? | Real sample | Linear range | Detection limit | Year published |
|---|---|---|---|---|---|---|---|
| MWCNTs-SPE | Nafion/Bi | Pb, Cd, Zn | Yes | Lake/Drinking/Tap water | 0.05-100 ppb<br>0.5-80 ppb<br>0.5-100 ppb | 0.01 ppb<br>0.3 ppb<br>0.1 ppb | 2013 [113] |
| CNF-SPE | GSH | Pb, Cd | Yes | - | 10.1-150.1 ppb<br>10.8-150.1 ppb | 3.0 ppb<br>3.2 ppb | 2016 [114] |
| CNTs-PTFE-flow through system | - | Cu | - | - | 0.064-635.5 ppb | 0.064 ppb | 2016 [115] |
| SWCNT-AuNPs | - | Pb, Cd | Yes | - | 3.31-22.29 ppb | 0.546 ppb<br>0.613 ppb | 2012 [116] |
| MWCNTs-GO-GC | - | Pb, Cd | Yes | Electroplating effluent | 0.5-30 ppb | 0.2 ppb<br>0.1 ppb | 2014 [117] |
| MWCNT-CPE | Bi | Cd | - | Tap water | 1-60 ppb | 0.3 ppb | 2013 [55] |
| MWCNT/$Fe_3O_4$ | DMF | Pb | - | Distilled/tap/river water | 0.004-0.33 ppb | 1.24 ppt | 2013 [118] |
| MWCNT/SbNPs-CPE | - | Pb, Cd | Yes | Wheat flour | 10-60 ppb | 0.65 ppb<br>0.77 ppb | 2014 [119] |
| MWCNT/GO/-GC | Nafion | Pb, Cd | Yes | Electroplating effluent | 0.5- 30 ppb | 0.2 ppb<br>0.1 ppb | 2014 [117] |
| GO/MWCNT-ITO | - | Pb, Cu | Yes | - | 0.05-2.5 µM<br>10.4-518.0 ppb<br>3.2-158.9 ppb | 1.2 ppb<br>0.76 ppb | 2015 [120] |

SPE: screen printed electrode; GSH: glutathione; PTFE: Polytetrafluoroethylene; GO: graphene oxide; CPE: carbon paste electrode; ITO: indium tin oxide

## 3. Analysis of real samples

Different from standard solutions containing only limited types of heavy metal ions, real samples as simple as tap water are much more complex than standard solutions, not to mention wastewater, river water and biological samples of blood, urine, and other types of body fluid which contain various small and large molecules that can interfere with sensor performance by fouling the surface. A reliable method to test a sensor under development is to compare the results with other detection techniques. And depending on the sample source, various pretreatment may apply to extract metal ions. An MWCNT paste electrode for Cd and Hg modified by chitosan and glutaraldehyde (both used to improve the absorptivity and selectively for metal ions) was examined with natural water, wastewater, urine and sediments by the standard addition method. Sediments from the river were dried and microwave digested in aqua regia and $H_2O_2$ to extract ions. The analysis results were compared with Thermospray Flame Furnace Atomic Absorption Spectroscopy (AA) and Inductively Coupled Plasma Optical Emission Spectrometry (ICP-OES). The sensor was not statistically different from the reference methods at the 95% confidence interval.[64] In another study of using Hg/MWCNTs graphite electrode for Pb and Cd determination, $Pb^{2+}$ and $Cd^{2+}$ were found at 1.7 ± 0.06 ppb and 0.14 ± 0.04 ppb in tap water.[43] This electrode was also used to determine $Pb^{2+}$ in human hair, in which the ions were extracted through intense acid and base treatment and $Pb^{2+}$ was determined as 11.2 ± 0.2 μg/L in acetate buffer that agreed well with AA (10.8 ± 0.3 μg/L).[43] Another study determined the Hg and Pb concentration in tuna fish, shrimp, tobacco, and human teeth with an MWCNTs/Schiff base carbon paste electrode. In order to extract the ions, tuna fish and Shrimp were treated with intensive acid and heat, tobacco sample was microwave digested in $HNO_3$ and $H_2O_2$, human teeth were crushed and ground after series of acid washes.The concentrations determined by this electrode were in good agreement with AA as well.[91] An SWCNT modified on boron-doped p-type silicon wafer electrode with metal extractant (bis(2,4,4-trimethylpentyl) phosphinic acid (PA/d) showed good sensitivity toward Cr. The practical application of this electrode was verified in an electroplating effluent collected from a Chrome plating industry and ground water with the reference method of ICP-AES.[121] The CNTs sensors discussed above have the detection limits for heavy metal ions well below the EPA requirement (in drinking water).[122] The satisfactory application of these CNTs sensors in real samples, especially as the results are in good agreement with other analytical techniques, demonstrates the applicability of the CNTs electrodes for heavy metal ion sensing. However, for more complex bio-samples, intensive pretreatment are required to extract the ions and minimize the fouling to the electrode.

## Conclusions

This review has addressed recent progress in the applications of CNTs based electrodes for heavy metals sensing by stripping voltammetry. As improved methods of fabrication for both CNTs and its chemical modifiers have been discovered and optimized, the sensitivity and versatility of CNT-based voltammetry analyses for heavy metals have increased dramatically. The positive characteristics of CNTs electrode materials and electrode modifiers are further improved and specialized through the use of surface modification on both the open ends and side walls. The modified electrodes pose sensitivity and detection limits to heavy metals comparable and sometimes even better than conventional Hg electrode but with negligible toxicity; the versatility of combining CNTs with other electrode types such as GC and Bi, and the continuous discoveries of modification chemicals have set new records for improving the electrodes' overall performance for detecting heavy metals. However, researchers are still trying to fully understand how the CNTs work and perform, and the challenges of applying CNTs based heavy metal sensors for real sample analysis still stimulate the studies of CNTs.